\documentclass[aps,prx,twocolumn,superscriptaddress,noeprint,longbibliography, nofootinbib]{revtex4-1}

\usepackage{graphicx}
\usepackage{bm}
\usepackage{physics}
\usepackage{xcolor}
\usepackage{enumitem}
\usepackage{amsmath, amssymb}
\usepackage[normalem]{ulem}
\usepackage{natbib}
\usepackage{comment}

\newcommand{\be}{\begin{equation} }
\newcommand{\ee}{\end{equation} }
\newcommand{\ba}{\begin{eqnarray} }
\newcommand{\ea}{\end{eqnarray} }

\newcommand{\bit}{\begin{itemize}}
\newcommand{\eit}{\end{itemize}}

\graphicspath{{Figures/}}

\usepackage[colorlinks=true]{hyperref}
\hypersetup{citecolor = blue}

\begin{document}
\title{Level statistics detect generalized symmetries}

\author{Nicholas O'Dea}
\affiliation{Department of Physics, Stanford University, Stanford, CA 94305, USA}

\begin{abstract}
Level statistics are a useful probe for detecting symmetries and distinguishing integrable and non-integrable systems. I show by way of several examples that level statistics detect the presence of generalized symmetries that go beyond conventional lattice symmetries and internal symmetries. I consider non-invertible symmetries through the example of Kramers-Wannier duality at an Ising critical point, symmetries with nonlocal generators through the example of a spin-$1$ anisotropic Heisenberg chain, and $q$-deformed symmetries through an example closely related to recent work on $q$-deformed SPT phases. In each case, conventional level statistics detect the generalized symmetries, and these symmetries must be resolved before seeing characteristic level repulsion in non-integrable systems. For the $q$-deformed symmetry, I discovered via level statistics a $q$-deformed generalization of inversion that is interesting in its own right and that may protect $q$-deformed SPT phases.
\end{abstract}

\maketitle

\noindent \textbf{\textit{Introduction--}} 
The study of symmetries is central to physics, from counting spectroscopic lines to describing symmetry broken phases to using the Lorentz symmetry group of spacetime to constrain relativistic field theories. \textit{Generalized} symmetries are playing an increasing role in modern condensed matter and high energy physics. These generalized symmetries go beyond the set of spacetime symmetries and conventional internal symmetries generated by a sum or integral of local operators, and examples abound. Topological phases can be understood through symmetry-breaking of higher-form symmetries (see \cite{mcgreevygeneralized2023, gomesintroduction2023} for reviews, and references therein). So-called ``non-invertible" symmetries enjoy a close relation to fusion rules in $(1+1)d$ conformal field theories and can induce generalized Lieb-Schultz-Mattis theorems forbidding unique gapped ground states \cite{mcgreevygeneralized2023, seibergnoninvertible2024}. Continuous symmetries generated by sums of nonlocal ``tailed" operators can control phase diagrams, such as the phase boundary between the gapless $XY$ phase and the Haldane phase of anisotropic Heisenberg spin-$1$ antiferromagnets (see \cite{alcarazcritical1992, kitzawaphase1996} for discussions of the phase diagram and \cite{atsuhiroSU22003} for discovery of the symmetry). Symmetries can be $q$-deformed (often called a ``quantum group symmetry")~\cite{qgroupbook}, a structure-preserving one-parameter modification of Lie symmetries that maintains important properties like integrability ~\cite{majidquasitriangular1990}, quantum scars~\cite{odeatunnels2020}, and symmetry-protected topological (SPT) order~\cite{quellaaklt2020, quellabilinear2021, frankeduality2024} while fundamentally twisting the structure of the Lie algebra.

Symmetries are also key to the study of integrability, quantum chaos, and transport. Integrability or exact solvability of many-body Hamiltonians requires an extensive set of symmetries, often from free-particle methods or variants of the Bethe ansatz~\cite{franchiniintegrable2017, faddeevalgebraic1996}. To distinguish a chaotic, non-integrable Hamiltonian from an integrable one using conventional techniques like the study of level repulsion requires ``resolving" symmetry sectors of conventional symmetries or, in other words, only comparing the energy eigenvalues of states with the same symmetry quantum numbers. The eigenstate thermalization hypothesis posits a simple ansatz for matrix elements of local operators in the eigenbasis of a chaotic Hamiltonian, but the presence of symmetries induces correlations between matrix elements and requires an improved ansatz~\cite{murthynonabelian2023}. Mazur bounds~\cite{mazurnonergodicity1969} tie autocorrelation functions to conserved quantities and are especially important for describing ballistic transport; discrepancies in transport versus Mazur bound predictions led to the discovery of quasilocal conserved quantities in the integrable spin-$1/2$ $XXZ$ anisotropic Heisenberg spin chains~\cite{ProsenXXZ2016}.

Exactly what counts as a symmetry is subtle. Symmetries are often taken to include lattice symmetries and symmetries generated by sums of strictly local operators. In particular, rank-1 projectors onto eigenstates of a Hamiltonian are typically \textit{not} viewed as symmetries\footnote{There is an alternative perspective; works by Moudgalya and Motrunich on commutant algebras~\cite{moudgalyacommutant2022mar,moudgalyacommutant2023annals,moudgalyaexhaustive2209, moudgalyanumerical2023, moudgalyasymmetries2023} treat nonlocal operators including eigenstate projectors as symmetries of (families of) Hamiltonians. The additional requirement in this context is that operators must commute with every term of the Hamiltonian separately, not just with the Hamiltonian.}: they generically do not restrict thermalization, they are generically extremely nonlocal, and every Hamiltonian including otherwise non-integrable ones would then have a complete set of commuting symmetries. 

In this work, I will consider various examples of generalized symmetries that are neither lattice symmetries nor generated by sums of strictly local operators, but that nevertheless maintain more structure than eigenstate projectors. In the cases I consider, this extra structure includes compact representations of the symmetries or their generators by quantum circuits and/or matrix product operators. 

Though generalized symmetries are described by intricate and varied constructions, I argue that they can be readily detected through the distribution of a conventional but sensitive function of the energy eigenvalues: the $\tilde{r}$ ratio of consecutive gaps introduced by Oganesyan and Huse~\cite{Oganesyan2007} and further studied by Atas et al.~\cite{Atas2013}.
\begin{equation}\label{eq:tilder}
    \tilde{r}_n = \min\left (\frac{E_{n+2}-E_{n+1}}{E_{n+1}-E_{n}}, \frac{E_{n+1}-E_{n}}{E_{n+2}-E_{n+1}} \right)
\end{equation}
The distribution of $\tilde{r}$ over a given many-body system's spectrum conforms closely to simple Wigner surmises that distinguish exactly-solvable integrable and chaotic non-integrable systems~\cite{Atas2013}. However, only when all conventional symmetries (including lattice symmetries and internal symmetries generated by sums of local, few-body commuting terms) have been resolved will the level statistics show these characteristic distributions. In particular, the characteristic level repulsion of the non-integrable systems is washed out when consecutive gaps are calculated using eigenvalues from more than one symmetry sector. 


In this letter, I show that level statistics are even more sensitive. They further detect generalized symmetries, including noninvertible symmetries from Kramers-Wannier duality, non-abelian symmetries with generators with nonlocal tails, and $q$-deformed continuous and discrete symmetries. The paper is organized around numerically-investigated examples of each of these three kinds of generalized symmetries, including a case where such a symmetry was discovered without a priori knowledge: a $q$-deformed discrete inversion symmetry that may be interesting in its own right. 

These examples further show that these unusual symmetries must be resolved to use level statistics to argue a system is integrable or non-integrable. Distinguishing integrable and non-integrable systems is important for understanding dynamics and thermalization~\cite{vidmargge2016} or arguing that certain outlier states are quantum scars~\cite{serbynscarsreview2021, moudgalyascarsreview2022, anushyascarsreview2023} in otherwise non-integrable symmetry sectors. Each of the non-integrable Hamiltonians considered in this letter show clean level statistics described by the Wigner-Dyson GOE class, but only after completely resolving both conventional and generalized symmetries. In the supplemental~\cite{supplementary}, I show that resolving these symmetries allows me to recognize that a model previously identified as integrable is in fact non-integrable. 

\noindent \textbf{\textit{Non-invertible symmetries--}} In this section, I focus on the simplest example of a non-invertible symmetry. Kramers-Wannier duality is a useful tool for locating phase transitions in classical~\cite{kramerswannier1941} and quantum models~\cite{kogutlattice1979}, and at self-dual points falls into the class of non-invertible symmetries~\cite{mcgreevygeneralized2023, seibergnoninvertible2024}. For example, in the transverse-field Ising model in periodic boundary conditions, $H=-\sum_i \sigma^z_i \sigma^z_{i+1} - h \sum_i \sigma^x_i$, a duality of the form $\sigma^z_i \sigma^z_{i+1} \leftrightarrow \sigma^x_i$ shows that, given the assumption there is only one phase transition at positive $h$, that phase transition must occur at the self-dual point $h=1$. In particular, this duality exchanges the paramagnetic and ferromagnetic phases while the critical point is self-dual and left unchanged.

At the self-dual point, this Kramers-Wannier duality becomes a symmetry of the Hamiltonian. Note that there is no unitary transformation that can take $\sigma^z_i \sigma^z_{i+1} \leftrightarrow \sigma^x_i$ everywhere in the Hilbert space; importantly, $\prod_i \sigma^z_i \sigma^z_{i+1} = 1$, while $\prod_i \sigma^x_i$ has eigenvalues $1$ and $-1$, so $\sigma^z_i \sigma^z_{i+1}$ and $\sigma^x_i$ cannot be related for all $i$ by a similarity transformation. However, the Kramers-Wannier duality can be realized as a unitary symmetry strictly within the Ising spinflip $F \equiv \prod_i \sigma^x_i=1$ symmetry sector; alternatively, it can be written as a non-invertible operator on the whole Hilbert space in the form of a unitary times the projection operator $\frac{1+\prod_i \sigma^x_i}{2}$. See section 3 of the lectures by Shao~\cite{shaodone2024} for a few explicit realizations of the Kramers-Wannier symmetry. 

\begin{figure}[h]
\includegraphics[width=\linewidth]{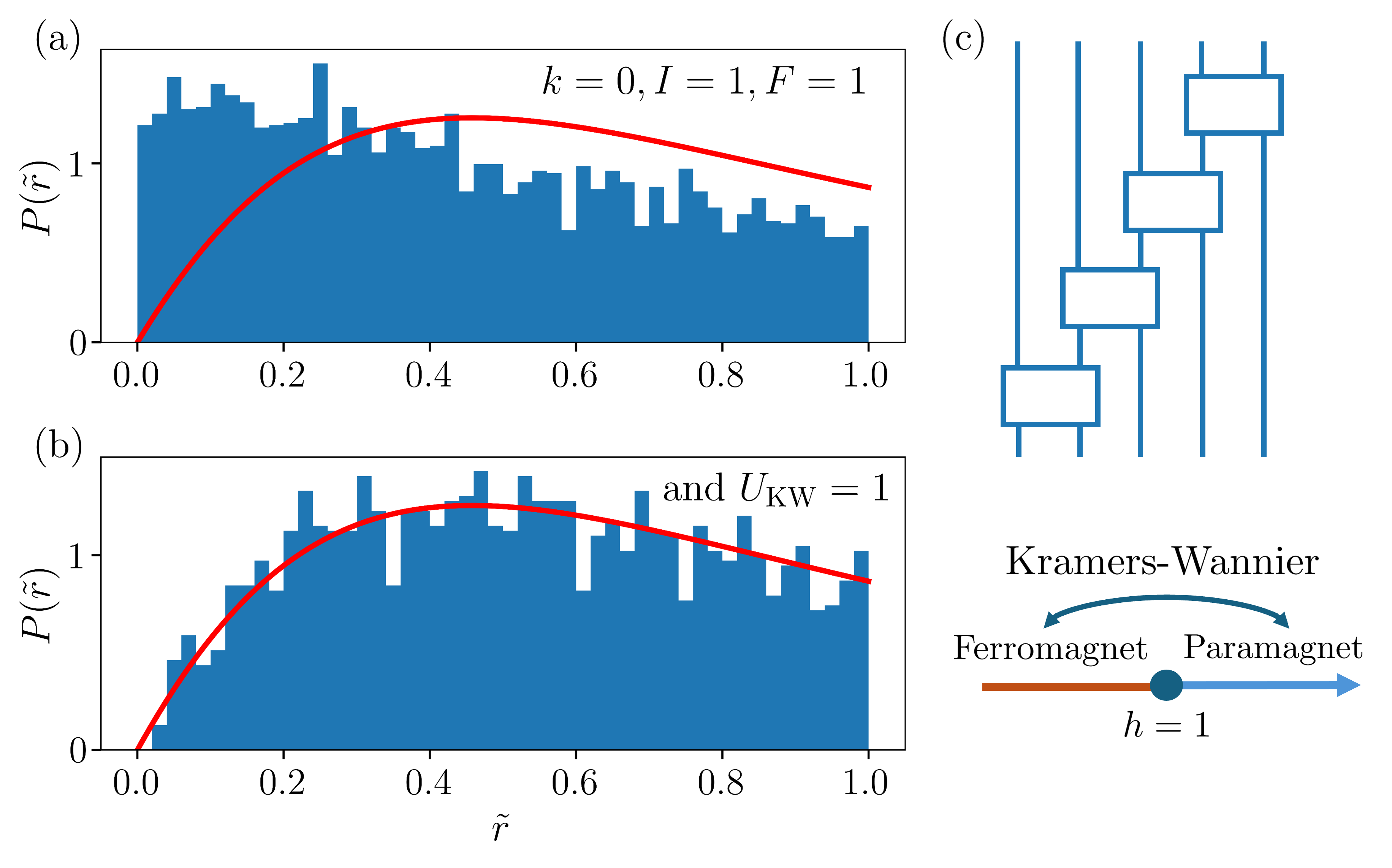}
 \caption{Level statistics for Hamiltonian in Eq.~\ref{eq:ANNNI} with parameters $h=1, a=.3$ and system size $L=18$. (a) Histogram of $\tilde{r}$ (Eq.~\ref{eq:tilder}) within the symmetry sector with momentum $k=0$, lattice inversion $I = 1$, Ising spinflip symmetry $F \equiv \prod_i \sigma^x_i = 1$, but missing Kramers-Wannier symmetry. Red line is Wigner's surmise for GOE~\cite{Atas2013}.
 (b) $\tilde{r}$ histogram within symmetry sector with same quantum numbers as (a) and with Kramers-Wannier quantum number $U_{\text{KW}}=1$. (c) Example ladder circuit construction of $U_{\text{KW}}$, with individual two-qubit gates corresponding to Eq.~\ref{eq:laddertwoqubit}, above a schematic showing the Kramers-Wannier duality mapping ferromagnetic and paramagnetic phases to one another.}
\label{fig:main1}
\end{figure}

I schematically depict the ladder circuit construction of~\cite{chensequential2024} in Fig.~\ref{fig:main1}c built out of the two-qubit operators
\begin{equation}\label{eq:laddertwoqubit}
    U_{i,i+1} = e^{-i\frac{\pi}{4} \sigma^x_{i+1}} e^{-i\frac{\pi}{4} \sigma^z_{i} \sigma^z_{i+1}}.
\end{equation} This ladder structure is not accidental. To convert a trivial ground state of a paramagnet to a ground state of a ferromagnet necessarily requires a linear-depth quantum circuit (i.e. number of layers going as $L$) to generate the quantum correlations of the latter~\cite{bravyicorrelations2006}. A ladder circuit is the ``most efficient" way to reach linear-depth and is part of the natural family of sequential circuits~\cite{chensequential2024} that map between phases more generally.

Note that the transverse-field Ising model is special; it is Jordan-Wigner transformable to free fermions and hence integrable~\cite{franchiniintegrable2017}. To test whether the Kramers-Wannier non-invertible symmetry is relevant for level statistics, I instead consider a common non-integrable deformation that preserves duality. This Hamiltonian
\begin{equation}\label{eq:ANNNI}
H = - \sum_i (\sigma^z_i \sigma^z_{i+1} + a \sigma^z_i \sigma^z_{i+2})  - h \sum_i (\sigma^x_i + a \sigma^x_i \sigma^x_{i+1})
\end{equation}
has a rich phase diagram and is self-dual at $h=1$~\cite{rahmanimajorana2015, milstedconformal2017}. In Fig.~\ref{fig:main1}, I show the distribution of $\tilde{r}$ with and without resolving the Kramers-Wannier duality symmetry. The difference is visible, with the characteristic level repulsion corresponding to a vanishing distribution at $\tilde{r}=0$ and agreement with the GOE Wigner's surmise~\cite{Atas2013} only appearing after restricting to a definite Kramers-Wannier symmetry sector. 

In the Supplemental Material~\cite{supplementary} Section I, I consider another duality, $\sigma^x_i \leftrightarrow \sigma^z_{i-1}\sigma^z_{i+1}$, which yields a non-invertible symmetry described by the fusion category $\text{Rep(D$_8$)}$. This symmetry is relevant to the study of non-invertible SPT phases,~\cite{seifnashricluster2024} and I show that, like the Kramers-Wannier symmetry discussed in this section, it is detected via level statistics.

\noindent \textbf{\textit{Continuous symmetries generated by sums of tailed operators--}}
I consider in this section continuous-symmetry generators with ``tails." By tails, I mean that a single-site term in a generator $G$, some $g_i$, now has products of operators extending to the left and/or right of $i$: 
\begin{equation}
G = \sum_i \left( \prod_{j <i} l_j \right) g_i \left( \prod_{k>i} r_k \right).
\end{equation} The resulting generalized symmetries cannot be expressed as an exponential of a sum of strictly local operators but nevertheless have important physical consequences and a direct effect on level statistics.

In particular, a nonlocal symmetry controls the phase boundary between the gapless $XY$ phase and the gapped Haldane phase in the spin-$1$ anisotropic Heisenberg model $H = \sum_i S^x_i S^x_{i+1} + S^y_i S^y_{i+1} + \Delta S^z_i S^z_{i+1}$. At the $XY$ point of $\Delta = 0$, there is an $SU(2)$ symmetry:
\begin{equation}\label{eq:tailedcommrel}
\begin{split}
    &Q^+ = \frac{1}{2}\sum_i \left( \prod_{j<i} e^{i \pi S^z_j} \right) (S^+_i)^2, \,\,\,\,\,\,\, Q^- = (Q^+)^{\dagger} 
    \\& Q^z = \frac{1}{2}[Q^+, Q^-],   \,\,\,\,\,\, Q^2 = \frac{1}{2}(Q^+ Q^- + Q^- Q^+) + (Q^z)^2 
\end{split}
\end{equation}
This symmetry ensures a degeneracy between the lowest excited states of the Haldane phase and the $XY$ phase, pinning the phase transition at $\Delta = 0$~\cite{atsuhiroSU22003}. This symmetry also plays a role in realizing scarred models; the states $(Q^+)^n|--..-\rangle$ are maximal-Casimir eigenstates of this $SU(2)$ symmetry, making them candidates for symmetry-based frameworks for embedding quantum scars (see Section 3 of the review~\cite{moudgalyascarsreview2022}; these states were first studied as quantum scars in~\cite{ISXY})). 

Here, the ``tails" of $Q^\pm$ extend to the left and are built out of $l_j = e^{i \pi S^z_j}$ which commutes with $(S^+)^2_i$ for all $i,j$. By a general result for tailed operators in Appendix A of Ref.~\cite{odeatunnels2020}, $Q^\pm$ has a compact, finite-bond-dimension representation as an MPO (which in turn induces a finite-bond-dimension representation of $Q^2$). The same result holds for the similarly-tailed $q$-deformed operators $\tilde{S}^\pm$ discussed in the following section. Thus, despite being sums of nonlocal operators, the tailed generators considered in this work share several features with conventional generators, including restricted entangling power and finite-bond-dimension representations as MPO's. 

\begin{figure}[h]
\includegraphics[width=\linewidth]{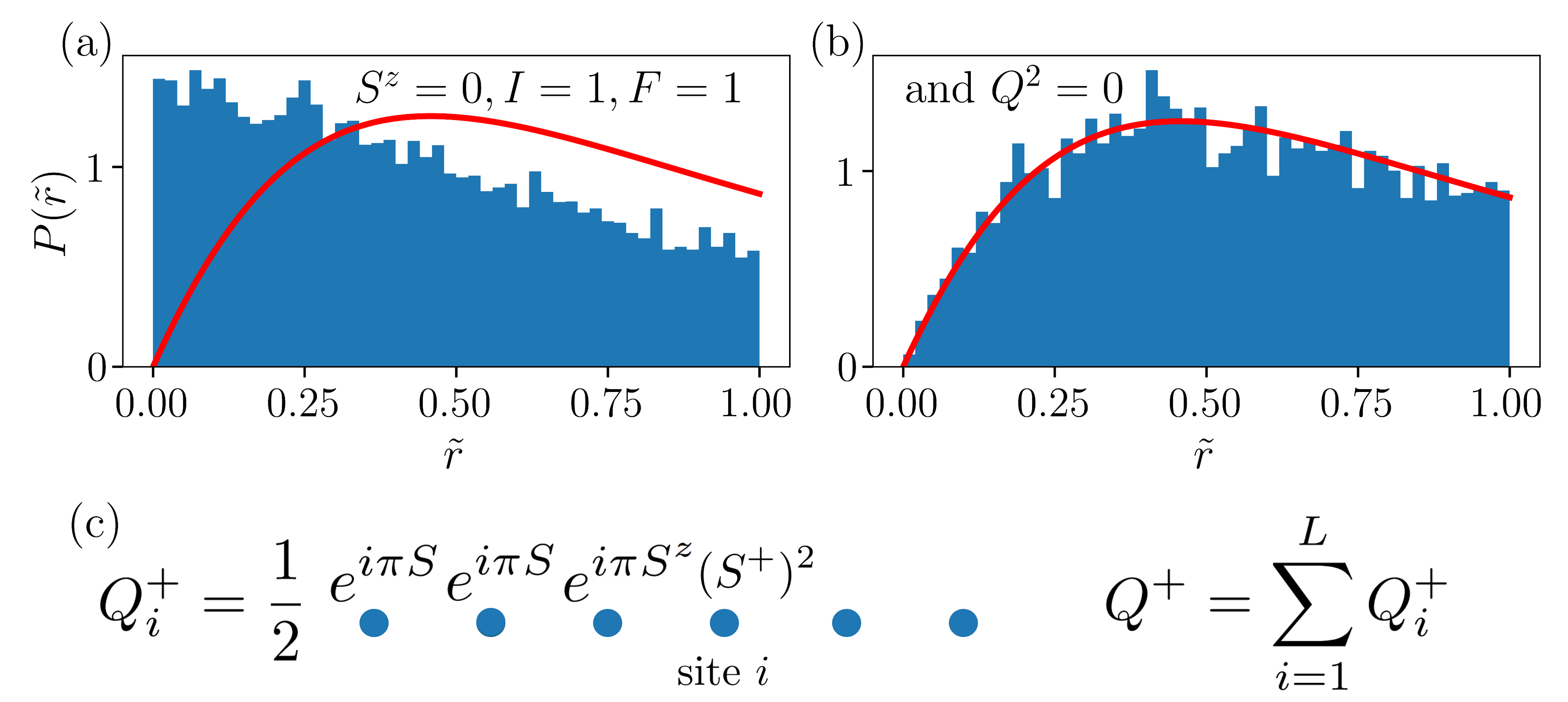}
 \caption{Level statistics for $XY$ Hamiltonian in Eq.~\ref{eq:XY} for system size $L=12$. (a) Histogram of $\tilde{r}$ (Eq.~\ref{eq:tilder}) within the symmetry sector with total $S^z=0$, lattice inversion $I = 1$, spinflip symmetry $F\equiv \bigotimes_i \begin{pmatrix} 0 & 0 & 1 \\ 0 & 1 & 0 \\ 1 & 0 & 0 \end{pmatrix}_i =  1$, but missing the Casimir $Q^2$. Red line is Wigner's surmise for GOE~\cite{Atas2013}.
 (b) $\tilde{r}$ histogram within symmetry sector with same quantum numbers as (a) and with Casimir $Q^2=0$. (c) Depiction of individual term of $Q^+$ showing tail of $\prod_{j<i} e^{i \pi S^z_j}$ extending to left of $(S^+_i)^2$}
\label{fig:main2}
\end{figure}

From the definition in Eq.~\ref{eq:tailedcommrel}, $Q^z = \frac{1}{2} \sum_{i=1}^{L} S^z_i $ is proportional to the usual $z$-component of total spin $S^z$. On the other hand, the Casimir $Q^2$ gives an extra quantum number that directly affects the level statistics. I show in Fig.~\ref{fig:main2} the histogram of $\tilde{r}$ for a symmetry sector of the spin-$1$ $XY$ Hamiltonian 
\begin{equation}\label{eq:XY}
    H_{XY} = \sum_{i=1}^{L-1} S^x_i S^x_{i+1} + S^y_i S^y_{i+1}.
\end{equation}
The level repulsion at $\tilde{r}$ is only manifest after further projecting into a sector with a unique quantum number for $Q^2$. 

As noted in the introduction, resolving all symmetries is necessary before determining whether a system is integrable or nonintegrable. In the Supplemental Material~\cite{supplementary} Section II, I note that a generalization of the $XY$ Hamiltonian (including Eq.~\ref{eq:XY} as a special case) was incorrectly deemed integrable in the literature -- the existence of the nonlocal $SU(2)$ symmetry was recognized, but the $Q^2$ Casimir was not resolved. 

\noindent \textbf{\textit{$q$-deformed symmetries--}} 
$q$-deformations are structure-preserving deformations that occur throughout mathematics. Many quantities, from binomial coefficients to derivatives to Lie algebras, have $q$-deformed analogues. All of these centrally feature the definition of a $q$-deformed complex number $[x]_q$
\begin{equation}
    [x]_q = \frac{q^x - q^{-x}}{q-q^{-1}}.
\end{equation}
Note that $\lim_{q \to 1} [x]_q = x$, so a $q$ deformation returns to the original structure on taking $q$ to $1$. For a mathematical introduction to $q$-deformations, see Ref.~\cite{qgroupbook}. For introductions with a physics focus, see Ref.~\cite{saleurintegrable1990} on integrability and Ref.~\cite{quellaaklt2020} on $q$-deformed SPT phases. I summarize a few of the key properties of $q$-deformations of $SU(2)$ here, which I'll call $SU_q(2)$. A $q$-deformation of a Lie group symmetry can be viewed as changing the ``coproduct" that describes how to add representations. For example, in a usual $SU(2)$ system, one can write the total raising operator for a couple spins as $S^+_{tot} = S^+ \otimes I + I \otimes S^+$. A $q$-deformation is such that the tensor product of representations requires tails:
\begin{equation}
    \tilde{S}^+_{tot} = \tilde{S}^+ \otimes q^{\tilde{S}^z} + q^{-\tilde{S}^z} \otimes \tilde{S}^+
\end{equation}
To realize this coproduct, the generators of the $q$-deformed Lie group (which no longer exponentiate to a Lie group) must obey new commutation relations
\begin{equation}\label{eq:qcomm}
    [\Tilde{S}^z, \Tilde{S}^\pm] = \pm \Tilde{S}^\pm\text{ and }[\Tilde{S}^+, \Tilde{S}^-] = [2\Tilde{S}^z]_q
\end{equation}
where 
\begin{equation}
    [x]_q = \frac{q^x - q^{-x}}{q-q^{-1}}
\end{equation}
is the $q$ deformation of a quantity $x$. Note that $\lim_{q \to 1} [x]_q = x$, which makes the $q \to 1$ limit of Eq.~\ref{eq:qcomm} recover the familiar commutation relations of $SU(2)$. In the following, I will consider generic positive $q>0$.
These new commutation relations also change the single-site representation of $q$-deformed spin $S$ operators for $S>1$, though mildly: $\langle m' | (\Tilde{S}^\pm) |m\rangle = \sqrt{[S\mp m]_q [S\pm m+1]_q }\delta_{m', m\pm 1}$. $\tilde{S}^z$ remains equal to $S^z$.

Important structures, including integrability~\cite{majidquasitriangular1990}, are maintained by $q$-deformations. The spin-$1/2$ $XXZ$ model can be meaningfully viewed as a $q$-deformation of the isotropic Heisenberg model, with $SU_q(2)$ symmetry inducing degeneracies\footnote{Because of the tails in $\tilde{S}^+$, $SU_q(2)$ symmetry requires special boundary conditions; open boundary conditions require boundary fields, and closed boundary conditions require a twist relative to the bulk interactions.} in the $XXZ$ model much like the $SU(2)$ symmetry of the isotropic Heisenberg model induces degeneracies. $q$-deformed groups furnish solutions of the Yang-Baxter equation (the so-called $\mathcal{R}$-matrix), making them key to the structure of integrability in interacting integrable systems. In Ref.~\cite{odeatunnels2020}, collaborators and I found that $q$-deformations can preserve quantum scarring -- certain perturbations to the spin-$1$ XY and AKLT models that would naively destroy quantum scars as eigenstates instead merely deformed the scar states instead of destroying them completely. It has been noted by Thomas Quella that important features of symmetry protected topological (SPT) phases are preserved in $q$-deformed models and that these features are potentially protected by $q \leftrightarrow 1/q$ dualities~\cite{quellaaklt2020, quellabilinear2021, frankeduality2024}. 


$q$-deformation shares some similarities and differences with the nonlocally-generated symmetry of the spin-$1$ $XY$ model in the previous section. Both symmetries have tails, though the tails in the $XY$ model's symmetry are perhaps more gentle -- the tails are made of products of unitary operators, while the $q$-deformation's tails are generically non-unitary. The $q$-deformation explicitly changes the relations that the algebra of generators must satisfy (see Eq.~\ref{eq:qcomm}), while the symmetry of the $XY$ model satisfies the commutation relations of vanilla $SU(2)$.

In the Supplemental Material~\cite{supplementary} Section III, I show that the Casimir of $SU_q(2)$,
\begin{equation}
    \tilde{S}^2 = \tilde{S}^-\tilde{S}^+ + [S^z]_q[S^z+1]_q
\end{equation}
has a direct effect on level statistics of $q$-deformed models. However, I want to instead emphasize in this section the existence of $q$-deformed versions of discrete symmetries that can also be detected through level statistics. 

\begin{figure}[h]
\includegraphics[width=\linewidth]{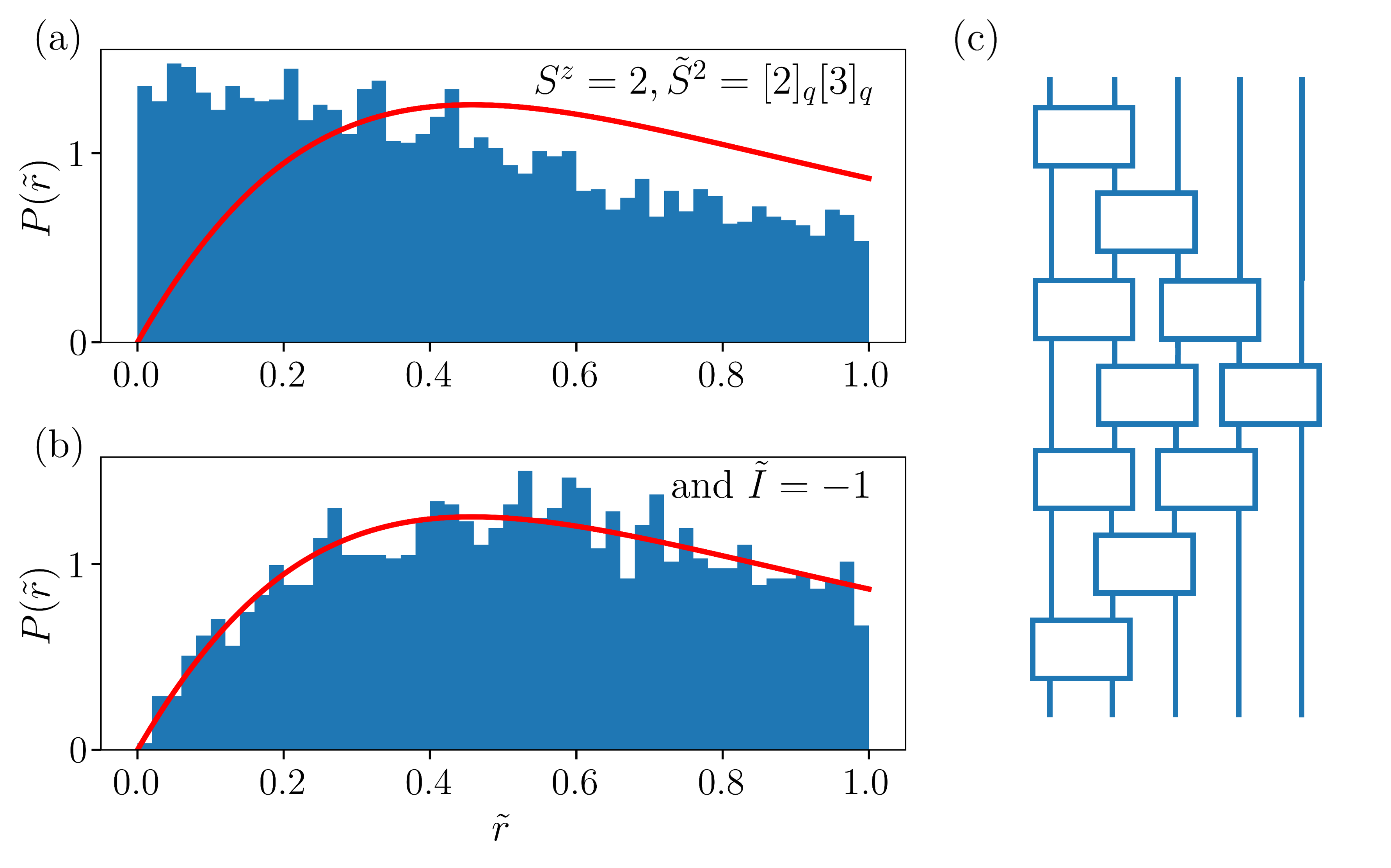}
 \caption{Level statistics for $q$-deformed AKLT Hamiltonian in Eq.~\ref{eq:AKLT} with deformation parameter $q=1.2$ at system size $L=11$. (a) Histogram of $\tilde{r}$ (Eq.~\ref{eq:tilder}) within the symmetry sector with $S^z=0$, $\tilde{S}^2 = [2]_q [3]_q$. Red line is Wigner's surmise for GOE~\cite{Atas2013}.
 (b) $\tilde{r}$ histogram within symmetry sector with same quantum numbers as (a) and with $q$-deformed inversion quantum number $\tilde{I}=-1$. (c) Circuit construction of $\tilde{I}$, with individual two-qutrit gates corresponding to $\check{\mathcal{R}}$ matrix ($q$-deformed SWAP).}
\label{fig:main3}
\end{figure}

In Fig.~\ref{fig:main3}, I consider the $\tilde{r}$ histogram of the $q$-deformed spin-$1$ AKLT model~\cite{Batchelorqdef, Klumperqdef, Totsuka1994, Motegi2010, Santos2012}
\begin{equation}\label{eq:AKLT}
    H = \sum_{i=1}^{L-1} \tilde{P}^{(2)}_{i,i+1}
\end{equation}
where $\tilde{P}^{(2)}_{i,i+1}$ is the projector onto two-site total $q$-deformed spin $2$; i.e. a projector onto two-site $\tilde{S}^2 = [2]_q [3]_q$. The explicit form of this Hamiltonian in terms of the usual spin-$1$ $\vec{S}$ is given in ~\cite{quellaaklt2020}, where it was studied as a prototypical $q$-deformed SPT that lacks the usual protecting symmetries~\cite{pollmannentanglement2010} like inversion. This Hamiltonian is symmetric under $SU_q(2)$ (technically $SO_q(3)$).

Even on resolving total $\tilde{S}^2$, an additional symmetry of $\tilde{I}$ must be resolved before the level statistics show level repulsion and follow the characteristic GOE Wigner's surmise. $\tilde{I}$ is a $q$-deformed inversion symmetry that is, to my knowledge, novel. I have given an empirical construction of this symmetry in Fig.~\ref{fig:main3}c). The idea of the construction is that spatial inversion can be realized through a sequence of SWAPs. A natural $q$-deformation of inversion follows if every SWAP gate is replaced by a $q$-deformed one: the $\check{\mathcal{R}}$ (``R-check'') matrix, which reduces to SWAP on $q \to 1$ and which I give explicitly in the Supplemental Material~\cite{supplementary} Section IV. The resulting operator is $\mathbb{Z}_2$ valued up to a prefactor that depends on total $\tilde{S}^2$; in the sector $\tilde{S}^2=[n]_q [n+1]_q$, $\tilde{I}$ takes on the values $\pm $ up to a constant multiplicative factor of $q^{n(n+1)-2}$.~\footnote{From numerical checks of $q$-deformed spin-$S$ chains, the $-2$ factor appears to be $-S(S+1)$.}

I discovered this symmetry because the level statistics of the $q$-deformed AKLT model in Fig.\ref{fig:main3}a) did not agree with my belief that the model should be nonintegrable. That is, this symmetry furnishes an example where level statistics were used to motivate the existence of a missing symmetry before its ultimately successful construction. 

$\tilde{I}$ is also interesting in its own right. Signatures (such as degeneracy in the entanglement spectrum) of the conventional spin-$1$ Haldane SPT phase persist under preserving only spatial inversion symmetry~\cite{pollmannentanglement2010}; the existence of $\tilde{I}$ suggests $q$-deformed SPT phases that are protected solely by the $\tilde{I}$ symmetry. 

\noindent \textbf{\textit{Discussion--}} 
I have shown through a few examples that level statistics successfully detect unusual symmetries, including Kramers-Wannier duality at the self-dual point, non-local SU(2), and $q$-deformed continuous and discrete symmetries. Level statistics thus provide a straightforward and sensitive tool to rapidly check for these unusual and generalized symmetries.

My work has exclusively focused on exact symmetries. Symmetries can also emerge at low energies and long distances -- for example, explicitly breaking higher-form symmetries often leads to a whole phase where there exists a low-energy subspace with an emergent version of the explicitly broken symmetry. At critical points that flow to CFTs, there are emergent symmetries in the ground state reflecting the enhanced symmetries of the renormalization group fixed point, and these can be captured through MPO techniques~\cite{mingrudetecting2023}. My level statistics numerics probed the entire spectrum, but a more precise study of the spectrum at low energy densities may be able to capture signatures of emergent symmetries as well. 

\textbf{\textit{Acknowledgements--}}
I thank the ARCS Foundation for ARCS Scholar funding under which this research was performed.
I thank Thomas Quella for useful discussions about whether duality transformations in $q$-deformed models can be lifted to symmetries. His works on $q$-deformed SPT phases served as my introduction to quantum groups. I also thank him for calling to my attention the $\text{Rep(D$_8$)}$ symmetry considered in Section I of ~\cite{supplementary}. I thank Wen Wei Ho for useful comments on Section II of ~\cite{supplementary}. I acknowledge Physics Stack Exchange users Gec, Meng Cheng, and Ruben Verresen for questions, comments, and answers related to Kramers-Wannier symmetries~\cite{PSE472064, PSE681849}; in particular, Gec's question on integrability at critical points prompted my answer, which ultimately became the seed of this paper. I thank Anushya Chandran and Fiona Burnell for past collaborations on quantum scars, including one~\cite{odeatunnels2020} that led me to realize via level statistics that there was a missing $q$-deformed inversion symmetry in the $q$-deformed AKLT model. I especially thank my advisor Vedika Khemani for past and current collaborations and encouraging me to publish these results.

\bibliography{main.bib}

\end{document}


\title{Supplemental Material for: Level statistics detect generalized symmetries}

\author{Nicholas O'Dea}
\affiliation{Department of Physics, Stanford University, Stanford, CA 94305, USA}
\maketitle
\renewcommand{\thefigure}{S\arabic{figure}}
\renewcommand{\theequation}{S\arabic{equation}}

Section I of this supplemental material gives an additional example of a non-invertible symmetry -- one protecting a symmetry-protected topological (SPT) phase that includes the cluster state -- and shows that resolving it is necessary to realize good level statistics. Section II discusses a family of models that included the spin-$1$ XY model. This family was identified as integrable in the literature, but I show that it is in fact non-integrable via a proper treatment of its symmetries. Sections III and IV discuss details of the $q$-deformed symmetries. Section III gives an example showing that the $q$-deformed $\tilde{S}^2$ must be resolved for good level statistics, while Section IV gives the explicit form of $\check{\mathcal{R}}$, the building block of the $q$-deformed inversion symmetry.

\section{Non-invertible symmetry protecting the cluster state SPT phase}
In the main text, I considered a non-invertible symmetry arising from a Kramers-Wannier duality, $\sigma^z_i \sigma^z_{i+1} \leftrightarrow \sigma^x_i$. In this appendix, I consider a duality of a similar flavor, $\sigma^z_{i-1} \sigma^z_{i+1} \leftrightarrow \sigma^x_i$, which was identified in Ref.~\cite{seifnashricluster2024} with a non-invertible symmetry ``$D$" of the cluster state~\cite{raussendorfmbqc2003, nielsencluster2006} Hamiltonian $-\sum_{i=1}^L \sigma^z_{i-1} \sigma^x_i \sigma^z_{i+1}$. 

The cluster state Hamiltonian is a paradigmatic example of a $\mathbb{Z}_2 \times \mathbb{Z}_2$ SPT phase~\cite{sontopological2011}. This $\mathbb{Z}_2 \times \mathbb{Z}_2$ is generated by $F_e \equiv \prod_{\text{even } i} \sigma^x_i$ and $F_o \equiv \prod_{\text{odd } i} \sigma^x_i$ spinflip symmetries on the even and odd sublattices. The cluster state Hamiltonian is additionally in a non-invertible SPT phase protected by $D$, which is itself described by the fusion category $\text{Rep(D$_8$)}$~\cite{seifnashricluster2024}. 

Note that $\sigma^z_{i-1} \sigma^z_{i+1} \leftrightarrow \sigma^x_i$ cannot be realized unitarily everywhere in the Hilbert space. In particular, under this duality, both $F_e$ and $F_o$ would become $1$ despite having both $-1$ and $1$ as eigenvalues. However, this duality can be realized as a unitary symmetry strictly within the sector $F_e = F_o = 1$; alternatively, it can be written as a non-invertible operator on the whole Hilbert space in the form of a unitary operator times the projector $\frac{1+F_o}{2} \frac{1+F_e}{2}$. In the following, I use the explicit form of the non-invertible symmetry given in Eq. 5 of Ref.~\cite{seifnashricluster2024}. 

The cluster state Hamiltonian is trivially integrable, as all of the terms commute with one another. I instead consider a non-integrable deformation of the cluster state Hamiltonian
\begin{equation}
\begin{split}\label{eq:defcluster}
    H = -\sum_{i=1}^L \big(  &(\sigma^z_{i-1} \sigma^x_i \sigma^z_{i+1}) + a(\sigma^x_i  + \sigma^z_{i-1} \sigma^z_{i+1})
    \\& + b(\sigma^x_i \sigma^x_{i+1} + \sigma^z_{i-1} \sigma^z_i \sigma^z_{i+1} \sigma^z_{i+2}) \big)
\end{split}
\end{equation}
that maintains the non-invertible symmetry $D$.

\begin{figure}[h]
\includegraphics[width=\linewidth]{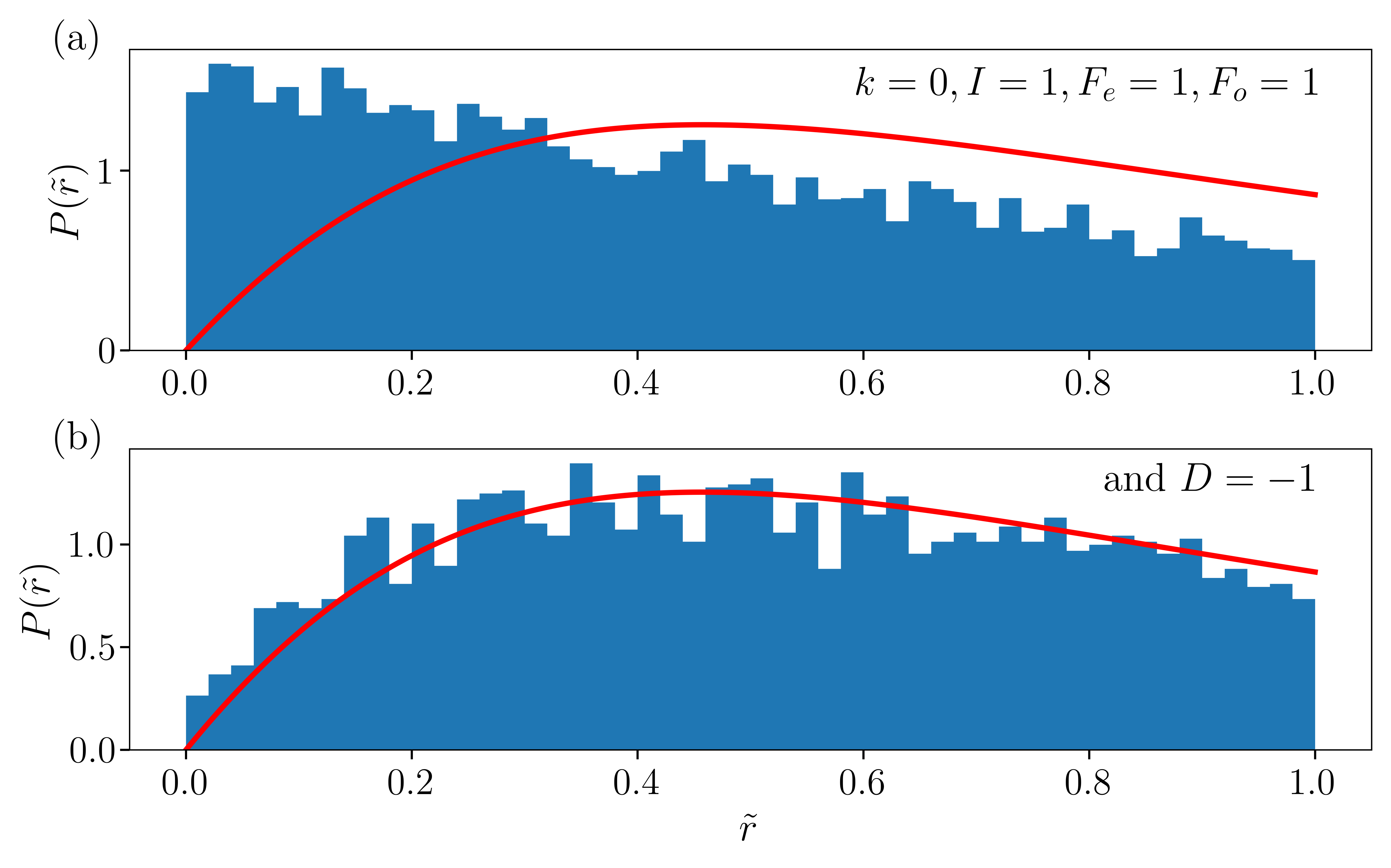}
 \caption{Level statistics for Hamiltonian in Eq.~\ref{eq:defcluster} for system size $L=20$, $a=.3$, and $b=.2$. (a) Histogram of $\tilde{r}$ (Eq.~1 of main text) within the symmetry sector with momentum $k=0$, lattice inversion $I = 1$, even-sublattice spinflip symmetry $F_e=1$, odd-sublattice spinflip symmetry $F_o=1$, but missing the non-invertible symmetry $D$ (Eq. 5 of Ref.~\cite{seifnashricluster2024}). Red line is Wigner's surmise for GOE~\cite{Atas2013}.
 (b) $\tilde{r}$ histogram within symmetry sector with same quantum numbers as (a) and including $D=-1$.}
\label{fig:app1}
\end{figure}

In Fig.~\ref{fig:app1}, I show that the level statistics within the $F_e = F_o =1$ sector (where the non-invertible symmetry is active) indeed only show level repulsion and follow the GOE Wigner's surmise after resolving the non-invertible symmetry $D$.

In passing, I note that in Fig.~\ref{fig:app1}, I chose to depict the level statistics from the symmetry sector $D=-1$. When I instead consider the symmetry sector $D=1$, I do not yet see level repulsion. Following the spirit of this work, I believe this signals an additional unresolved symmetry within the $D=1$ sector.

\section{Non-integrability of spin-$1$ $XY$ model}\label{app:nonintXY}
In the main text, I commented that the $Q^2$ $SU(2)$ symmetry of the spin-$1$ XY model has led to incorrect statements about its integrability. Ref.~\cite{Chattopadhyay} argued that the twisted XY Hamiltonian of the form
\begin{equation}
\begin{split}\label{eq:XYgeneral}
    H' &= \sum_{i=1}^{L-1} J_{i,i+1} (S^x_i S^x_{i+1} + S^y_i S^y_{i+1}) 
    \\&+ J_{L,1} \frac{1}{2}(S^+_L S^-_{1} e^{\mp i \frac{\pi}{2} S^z} + S^-_L S^+_{1} e^{\pm i \frac{\pi}{2} S^z})
\end{split}
\end{equation}
where $S^z$ is the total $z$-magnetization is integrable for arbitary real couplings. In particular, the authors write that for generic choices of $J_{i,i+1}$,

`` [...] and so [$H'$'s] energies and eigenstates
are organized in representations of this ``twisted" SU(2) algebra. In other words, this model, while interacting, is integrable. We therefore do not expect its level statistics, even upon resolving all possible global symmetries (magnetization, translation, if it exists, etc.), to tend towards a WD [Wigner-Dyson] class."

\begin{figure}[h]
\includegraphics[width=\linewidth]{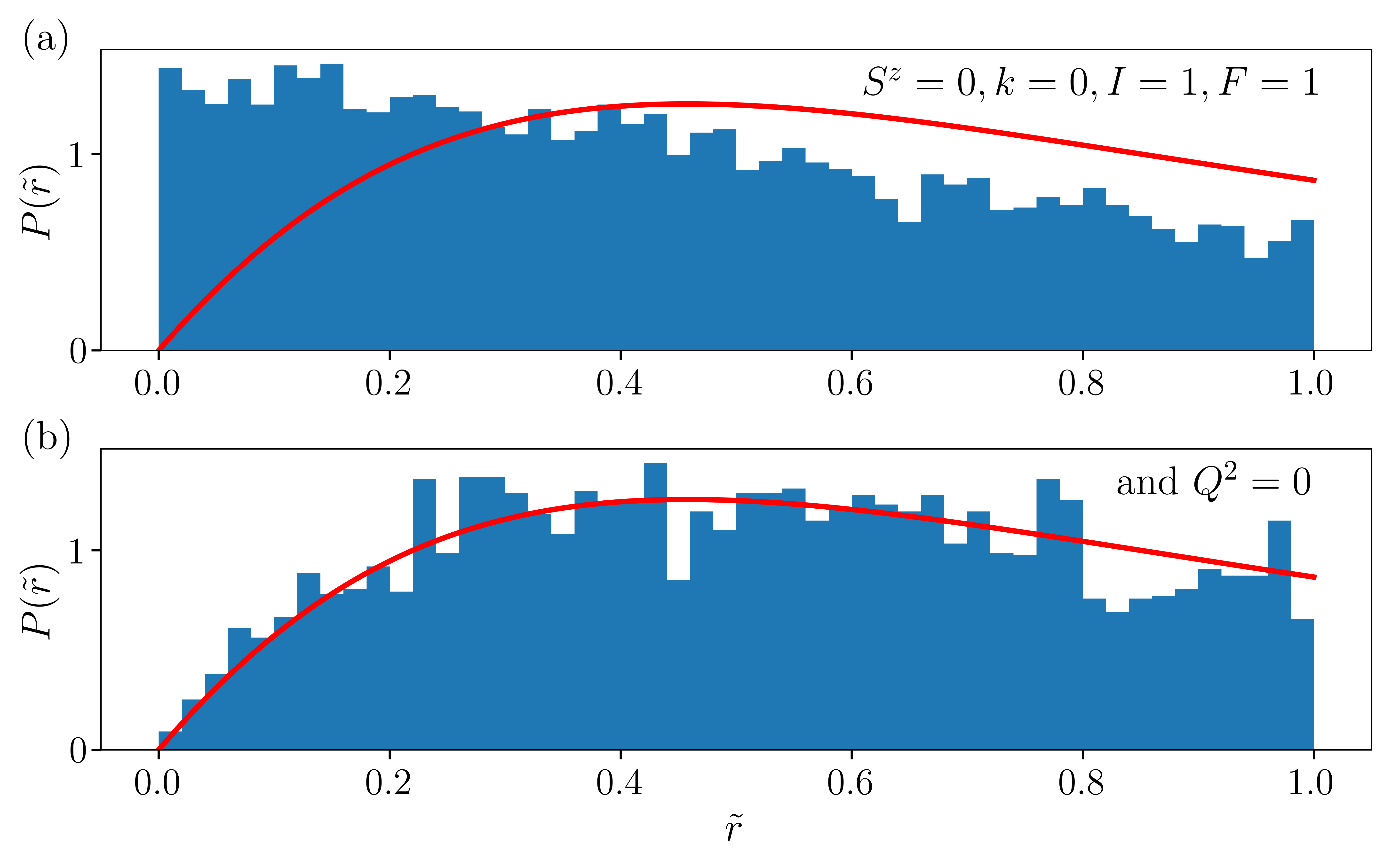}
 \caption{Level statistics for $XY$ Hamiltonian in Eq.~\ref{eq:XYgeneral} for system size $L=14$ and all $J_{i,i+1}=1$. (a) Histogram of $\tilde{r}$ (Eq.~1 of main text) within the symmetry sector with total $S^z=0$, momentum $k=0$, lattice inversion $I = 1$, spinflip symmetry $F= 1$, but missing the Casimir $Q^2$. Red line is Wigner's surmise for GOE~\cite{Atas2013}.
 (b) $\tilde{r}$ histogram within symmetry sector with same quantum numbers as (a) and with Casimir $Q^2=0$.}
\label{fig:app2}
\end{figure}

However, apart from trivial cases, Eq.~\ref{eq:XYgeneral} is generically nonintegrable. For example, I found in the main text that the case of open boundary conditions without disorder ($J_{i,i+1}=1$ except for $J_{L,1} = 0$) is not integrable, and its distribution of $\tilde{r}$ is well-described by the GOE Wigner-Dyson class. Ref.~\cite{Chattopadhyay} also had a focus on the case where all $J_{i,i+1}=1$, but in Fig.~\ref{fig:app2}, I show that on resolving all symmetries in the $S^z=0$ sector, including the $Q^2$ Casimir discussed in the main text, the level statistics again match those of the GOE Wigner-Dyson class. 

I emphasize that Ref.~\cite{Chattopadhyay}'s incorrect characterization of the twisted $XY$ model as integrable does not invalidate \textit{any} of its main results, which are interesting and worthwhile to read. Instead, I share this example to further emphasize how important it is to resolve all symmetries before using level statistics as an indicator of integrability. I briefly note that the fact that these models are generically non-integrable actually \textit{enhances} the results of Ref.~\cite{Chattopadhyay}, which described ``bond-bimagnon" eigenstates of perturbed $XY$ models in the context of oscillatory dynamics induced by scar states in otherwise chaotic systems. Ref.~\cite{Chattopadhyay} introduced perturbations to ensure these bond-bimagnon states counted as scar states in nonintegrable symmetry sectors; these perturbations are actually not necessary, so a wider class of models has these bond-bimagnon eigenstates as scar states.\footnote{Note that these bond-bimagnon states should not be confused with the bimagnon states studied in ~\cite{ISXY}. As noted there, the $Q^2$ symmetry must still be explicitly broken for the bimagnon states to count as scar states.}

\section{$q$-deformed Casimir $\tilde{S}^2$ affects level statistics}
In the main text, I emphasized the existence of $q$-deformed inversion $\tilde{I}$ symmetry and showed that it must be resolved to realize good level statistics. In this section of the supplemental, I show that the $q$-deformed Casimir $\tilde{S}^2$, which I resolved in Fig. 3 in the main text, must also be resolved to see good level statistics. To disentangle the effects of $\tilde{I}$ and $\tilde{S}^2$, I'll consider a staggered $q$-deformed spin-$1$ AKLT model
\begin{equation}\label{eq:staggaklt}
    H = \sum_{i=1}^{L-1} \tilde{P}^{(2)}_{i,i+1} (1 + (-1)^i \delta)
\end{equation}
which explicitly breaks $\tilde{I}$ for odd system sizes.

\begin{figure}[h]
\includegraphics[width=\linewidth]{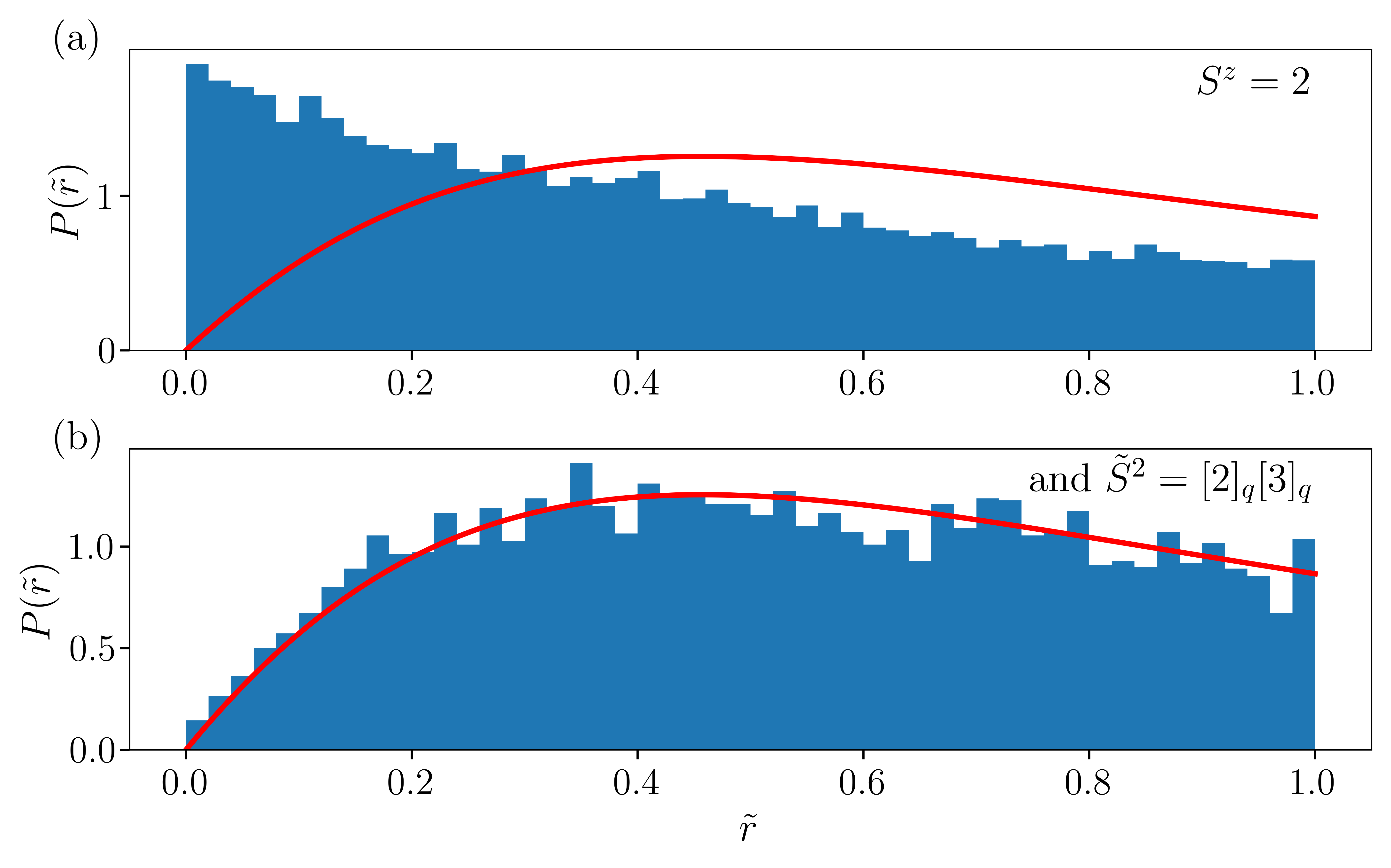}
 \caption{Level statistics for staggered AKLT Hamiltonian in Eq.~\ref{eq:staggaklt} for system size $L=11$. (a) Histogram of $\tilde{r}$ (Eq.~1 of main text) within the symmetry sector with total $S^z=2$ but missing the Casimir $\tilde{S}^2$ (Eq. in the main text). Red line is Wigner's surmise for GOE~\cite{Atas2013}.
 (b) $\tilde{r}$ histogram within symmetry sector with same quantum numbers as (a) and with Casimir $\tilde{S}^2$.}
\label{fig:app3}
\end{figure}

As anticipated, Fig.~\ref{fig:app3} shows that the level statistics are only well-described by the GOE Wigner-Dyson class when $\tilde{S}^2$ is taken into account. 

\section{Explicit form of $\check{\mathcal{R}}$}
There is an $\mathcal{R}$-matrix associated to $SU_q(2)$ which is defined to ``swap" the coproduct in the sense that
\begin{equation}
\mathcal{R}(\tilde{S}^+ \otimes q^{\tilde{S^z}} + q^{\tilde{-S^z}} \otimes \tilde{S}^+) \mathcal{R}^{-1} = q^{\tilde{S^z}} \otimes \tilde{S}^+ +  \tilde{S}^+ \otimes q^{\tilde{-S^z}}
\end{equation}
Such an $\mathcal{R}$-matrix satisfies the parameter-free Yang-Baxter equation. An explicit construction of the $\mathcal{R}$-matrix is given in Eq.~10 of Ref.~\cite{cramperevisiting2020} and rephrased here as
\begin{equation}\label{eq:rmatdef}
    \mathcal{R} = q^{\tilde{S}^z \otimes \tilde{S}^z} \sum_{n=0}^\infty \frac{(q-\frac{1}{q})^n}{[n]_q!} q^{n(n-1)/2}(\tilde{S}^+ q^{\tilde{S}^z} \otimes q^{-\tilde{S}^z} \tilde{S}^-)^n
\end{equation}
Here $[n]_q!$ is simply defined as $\prod_{k=1}^n [k]_q$. This equation naturally truncates; for example, $(\tilde{S}^+)^3=0$ for the $q$-deformed spin-$1$ representation considered in the main text, leaving only $n=0,1,2$ to contribute.

However, the key object of interest in the main text is the $\check{\mathcal{R}}$ matrix, which is simply defined as $\check{\mathcal{R}} = \text{SWAP } \mathcal{R}$. This object is a natural deformation of $\text{SWAP}$ that preserves the coproduct, and it is straightforward to see from Eq.~\ref{eq:rmatdef} that when $q\to 1$, then $\mathcal{R} \to I$ and hence $\check{\mathcal{R}} \to \text{SWAP}$. 

\begin{widetext}

For the $q$-deformed spin-$1$ representation, the explicit form of $\check{\mathcal{R}}$ is
\begin{equation}
    \check{\mathcal{R}} = 
    \begin{pmatrix} 
    q^2 & 0 & 0 & 0 & 0 & 0 & 0 & 0 & 0 \\
    0 & 0 & 0 & 1 & 0 & 0 & 0 & 0 & 0\\
    0 & 0 & 0 & 0 & 0 & \frac{1}{q^2} & 0 & 0& 0\\
    0 & 1 & 0 & q^2-\frac{1}{q^2} & 0 & 0 & 0 & 0& 0 \\
    0 & 0 & 0 & 0 & 1 & 0 & q-\frac{1}{q^3} & 0 & 0\\
    0 & 0 & 0 & 0 & 0 & 0 & 0 & 1 & 0\\
    0 & 0 &\frac{1}{q^2} & 0 & q-\frac{1}{q^3} & 0 & (q^2-\frac{1}{q^2})(1-\frac{1}{q^2}) & 0 & 0\\
    0 & 0 & 0 & 0 & 0 & 1 & 0 & q^2 - \frac{1}{q^2} & 0\\
    0 & 0 & 0 & 0 & 0 & 0 & 0 & 0 & q^2
    \end{pmatrix}
\end{equation}
This is the two-qutrit gate that builds up the $q$-deformed inversion symmetry in Fig.~3 in the main text.

\end{widetext}

\bibliography{main}